\documentclass[prl,twocolumn,superscriptaddress]{revtex4}

\newcommand{\calT}{{\cal T}}
\newcommand{\calF}{{\cal F}}
\newcommand{\calH}{{\cal H}}

\begin{document}

\title{Fractionalization, topological order, and
quasiparticle statistics}

\author{Masaki Oshikawa}
\affiliation{Department of Physics, Tokyo Institute of Technology \\
Oh-okayama, Meguro-ku, Tokyo 152-8551 Japan}

\author{T. Senthil}
\affiliation{Center for Condensed Matter Theory, Department of Physics, Indian Institute of Science \\
Bangalore 560 012, India}
\affiliation{
Department of Physics,\nolinebreak
Massachusetts Institute of Technology \\
Cambridge, Massachusetts 02139, USA}

\date{May 13, 2005}

\begin{abstract}
We argue, based on general principles,
that topological order is essential
to realize fractionalization in gapped insulating
phases in dimensions $d \geq 2$.
In $d=2$ with genus $g$,
we derive the existence of the minimum topological degeneracy $q^g$
if the charge is fractionalized in unit of $1/q$, irrespective
of microscopic model or of effective theory.
Furthermore, if the quasiparticle is either boson or fermion,
it must be at least $q^{2g}$.
\end{abstract}


\maketitle

\bigskip

Fractionalization of quantum numbers has been a focus of condensed
matter physics in recent years.
It refers to the emergence of a collective
excitation having fractional quantum numbers with
respect to the elementary particles (such as electrons),
in a strongly correlated system.
The notion of fractionalization is not only fascinating in itself,
but also has been related to other intriguing concepts in theoretical
physics as discussed in the following.

At present, several different systems
exhibit the
fractionalization~\cite{Laughlin-FQH,Kitaev,fraceft,SenthilFisher,Moessner,MSP,
BFG, Motrunich, Z3, PTinv}, at least theoretically. 
While the details naturally depend on each model under consideration,
the structure of the excitation spectrum is efficiently described
in terms of a gauge theory. 
More precisely the excitations consist of objects that
have long ranged non-local 
`statistical' interactions with each other which
may be encoded as an Aharanov-Bohm gauge interaction. 
This is well-known in the fractional quantum
hall effect where the fractionalized quasiparticles also
have fractional statistics.
Similarly in the fractionalized liquids described in 
Refs. ~\cite{fraceft,SenthilFisher,Moessner,MSP,BFG,Motrunich},
there are vison excitations that have long range 
statistical interactions with the fractionalized
particles (such as the spinons in a spin liquid). 

This emergent gauge structure generally implies the existence
of a certain kind of order
-- dubbed topological order -- associated with the global properties of 
the groundstate wavefunction~\cite{Wen}, 
which is also commonly found in the
above examples.
A characteristic signature of the topological order is
the groundstate degeneracy depending on the topology of the system.
This cannot be understood as a consequence of a conventional
spontaneous symmetry breaking, which is the standard mechanism
behind the groundstate degeneracy.
The intriguing nature and consequences of the topological order is best
understood in the Fractional Quantum Hall Liquid (FQHL)~\cite{Wen},
although some of the concepts are applicable to other systems.
In the gauge theory picture,
the topological degeneracy could be understood with different
``vacua'' corresponding to different number of vortices trapped
in each ``hole'' of the space (such as the torus.)


However, these developments apparently leave open the question on whether 
there is a different way to realize fractionalization
without the emergent gauge structures. 
As such, at this point it is also unclear whether
the topological order and associated groundstate degeneracy
are  necessary to have fractionalization.
As we are still far from the complete classification of
the fractionalized phases, and many novel examples of
fractionalization will likely be found in the future,
these questions would be of a significant importance.

In this paper, we demonstrate that there is indeed a general
and direct connection between the fractionalization
and the topological order,
in the specific context of systems with a fully gapped spectrum. 
Generalizing the gauge invariance argument
presented in Ref.~\cite{WHK},
the existence of the topological order is shown to follow
just from the fractionalization, irrespective of microscopic details.

As discussed before,
the known examples of the fractionalization are rather suggestive
of such a universal relation.
In the several examples of fractionalization
(at zero magnetic field) discussed
recently~\cite{fraceft,SenthilFisher,Moessner,MSP,BFG,Motrunich,Z3,PTinv},
the degeneracy is
at least $q^{2g}$-fold in a system on a $d=2$ dimensional
surface with genus $g$,
if the fractionalization occurs in the unit of $1/q$.
However, we must recall
that the Laughlin FQHL
does exhibit a groundstate degeneracy, but only
$q^g$-fold~\cite{TaoWu,WenNiu}. (See also Ref.~\cite{PTinv}.)
We will also clarify the difference between the cases with
$q^{2g}$-fold and $q^g$-fold degeneracy, which turns out to be
related to the statistics of the quasiparticles.
Not surprisingly, our argument is closely related to the
earlier studies on the topological order in FQHL
especially in Refs.~\cite{TaoWu,WenNiu}, and that in systems
of anyons~\cite{WDF}.

Now let us define the problem in a general setting.
We consider a system defined by a certain microscopic
Hamiltonian of interacting particles,
with an exact $U(1)$ global symmetry.
With the global $U(1)$ symmetry, we may assign a (fictitious) charge
to each particle, with the total charge being a conserved quantity.
We take the unit in which the elementary charge is unity, so that
the charge of all the particles appearing in the microscopic
model are integers.
We can now also introduce a (fictitious) external $U(1)$ gauge field
(``electromagnetic field'') coupled to the charge.
We set $\hbar=c=1$ so that the unit flux quantum
is given by $2\pi$.

The groundstate is generally a complicated state in terms
of the original particles.
Here we assume for simplicity that there is a finite gap above the
(possibly degenerate) groundstate(s).
We further assume that the elementary excitations of the
system are well-defined quasiparticles and quasiholes.
The quasiparticle may carry a charge that is a fraction
of the original unit charge, {\em thereby we define the
fractionalization.}
This definition is very natural and is independent of
the concrete model or mechanism of the fractionalization,
while it naturally applies to all the known cases.
Let us assume that the fractional charge of the quasiparticle
is $p/q$, where $p$ and $q$ are mutually prime integers.

For simplicity, let us consider a system on a $d=2$ torus
of sufficiently large size $L_x \times L_y$, for the moment.
We will comment on other cases later.
We define the following process as introduced
in Ref.~\cite{WenNiu}.
First we create a quasiparticle and its antiparticle (quasihole)
out of the vacuum (groundstate) at some location, and then
move the quasiparticle to $+x$ direction, so that it encircles
the torus to come back to the original location and to meet the
quasihole.
Finally, we pair-annihilate the quasiparticle and quasihole.
Here we assume that this process can be realized by a unitary
time evolution operator $\calT_x$ with respect
to a properly chosen time-dependent Hamiltonian,
e.g. with a time-dependent local
potential to create and drag the quasiparticle.
Thus we exclude quantum ``glassy'' systems as proposed
in Ref.~\cite{Chamon}.

Similarly, we introduce another unitary operator $\calT_y$,
corresponding to creation of a quasihole-quasiparticle pair and
annihilation after winding in $y$ direction.
It is expected that $\calT_{x,y}$
bring any state in the groundstate manifold state back,
at least approximately, to a (possibly different) groundstate.


Next we consider an adiabatic insertion of a unit flux quantum
$\Phi_0 = 2\pi$ through the ``hole'' of the torus, inducing an (fictitious)
electric field in $x$-direction.
This may again be realized by a time evolution in which the
$x$-component of the vector potential is gradually increased from $A_x=0$
to $A_x=2\pi/L_x$ in the Hamiltonian.
Thus it is represented by a unitary time-evolution operator ${\cal F}_x$.
We also define a similar operator ${\cal F}_y$ that corresponds to
an adiabatic insertion of a unit flux quantum through the other
``hole'', inducing the $y$-component of the vector potential. 
We assume that these operations do not close the gap to excitations above the groundstate manifold
and thus bring any groundstate to a  groundstate.
In $d=2$, it amounts to assuming the system to be an
insulator, while it is a stronger assumption
for $d \geq 3$.~\cite{Drude} (See also Ref.~\cite{PV}.)

Now let us consider two operations $\calT_x$ and $\calF_x$
in sequence. The flux insertion $\calF_x$ introduces the vector potential
$A_x=2\pi/L_x$, corresponding to the unit flux quantum $\Phi_0=2\pi$
contained in the system.
As we consider the process $\calT_x$ in different backgrounds,
let us distinguish them by denoting $\calT_x(\Phi)$
as the ``encircling'' process defined above
in the presence of the vector potential
$A_x=\Phi/L_x$.
The contained unit flux quantum does not induce the
Aharonov-Bohm effect on the original particles of integral charge.
However, for the quasiparticle with the fractional
charge $p/q$, the same vector potential $A_x=2\pi/L_x$
still gives a non-trivial Aharonov-Bohm phase
$e^{2 \pi i p/q}$ when the quasiparticle completes the
encircling process.
Thus we obtain a relation
\begin{equation}
 \calT_x(\Phi_0) \calF_x \sim e^{2\pi ip/q} \calF_x \calT_x(0).
\label{eq:tf1}
\end{equation}

On the other hand, because the microscopic model is given in terms of
the original particles of integral charge,
any Hamiltonian with an extra unit
flux quantum in the ``hole'' of the torus is unitary equivalent to the
Hamiltonian with zero flux.
Namely, the Hamiltonian $\calH(\Phi_0)$ with the unit flux quantum
is related to the Hamiltonian $\calH(0)$ with zero flux as
$ \calH(\Phi_0) = U^{-1} \calH(0) U$,
by a unitary operator $U$ which is called
as the large gauge transformation.
As we have argued previously, the encircling process $\calT_x$
should be realized by a time evolution with respect to an
appropriately chosen time-dependent Hamiltonian,
again {\em written in terms of the original particles}.
Therefore, the operator $\calT$ also must obey the relation
\begin{equation}
 \calT_x(\Phi_0) = U^{-1} \calT_x(0) U.
\label{eq:utu}
\end{equation}
Combining eqs.~(\ref{eq:tf1}) and~(\ref{eq:utu}), we obtain
\begin{equation}
 \calT_x(0) \tilde{\calF}_x \sim e^{2\pi i p/q} \tilde{\calF}_x \calT_x(0),
\label{eq:txfx}
\end{equation}
where $\tilde{\cal F}_x \equiv U \calF_x$.
In the following, for brevity $\calT_x$ without the argument denotes
$\calT_x(0)$, and likewise for $\calT_y$.

This algebra between $\calT_x$ and $\tilde{\calF}_x$ is identical to
that of the magnetic translation group, which we call as
the magnetic algebra.
By our assumptions, both of these
operators map a groundstate to a groundstate.
We thus immediately see that
the groundstates must be $q$-fold degenerate,
with the same reasoning as was used in Refs.~\cite{WenNiu,WDF}.
Our argument so far is essentially contained
in Ref.~\cite{WHK}, where the $q$-fold groundstate degeneracy of a
FQH liquid is derived based on the gauge invariance.
In this paper, we shall present a more systematic discussion
to demonstrate that the degeneracy is
topology-dependent, and that the degeneracy is also affected
by quasiparticle statistics.

For the other direction $y$, we obtain a corresponding relation
\begin{equation}
 \calT_y \tilde{\calF}_y \sim e^{2\pi i p/q} \tilde{\calF}_y \calT_y .
\label{eq:tyfy}
\end{equation}
Apparently, now we obtain two sets of the magnetic algebra, which would
imply a $q^2$-fold degeneracy on the torus.
However, as it should not apply to the Laughlin state
where the degeneracy is known to be only $q$-fold,
we have to examine more carefully
the interplay between eqs.~(\ref{eq:txfx}) and
(\ref{eq:tyfy}).

$\calF_x$ introduces the vector potential only in the $x$
direction, to which $\calT_y$ is insensitive.
Thus, with the large gauge transformation combined, we have
$ \tilde{\calF}_x \calT_y = \calT_y \tilde{\calF}_x$
and likewise for $\tilde{\calF}_y$ and $\calT_x$.
Therefore, we can take the basis in the groundstate subspace
so that $\tilde{\calF}_x$ and $\calT_y$ are both diagonalized.
Let the simultaneous eigenstate (among the groundstates) of them 
be $|f_x, t_y \rangle$ with $f_x$ and $t_y$ denoting the
eigenvalues of $\tilde{\calF}_x$ and $\calT_y$ respectively.
By applying $\calT_x$ to this state, one obtains
a new groundstate belonging to
a different eigenvalue $f_x e^{-2\pi i p/q}$
of $\tilde{\calF}_x$ because
\begin{equation} 
 \tilde{\calF}_x \big( \calT_x | f_x, t_y \rangle \big)
 = f_x e^{- 2\pi i p /q} \big( \calT_x | f_x, t_y \rangle \big)
\label{eq:fxchange}
\end{equation}
follows from eq.~(\ref{eq:txfx}).
By repeated applications of $\calT_x$, one can obtain
at least $q$ different groundstates as announced.

Similarly, we can apply $\tilde{\calF}_y$ to $|f_x , t_y \rangle$
to obtain $q$-fold degenerate groundstates belonging to different
eigenvalues of $\calT_y$.
The question now is whether these two procedures give different
set of groundstates.
It depends on whether (or how) the application of
$\calT_x$ changes the eigenvalue of $\calT_y$.
This boils down to the commutation relation between
$\calT_x$ and $\calT_y$, which actually
reflects the statistics of the quasiparticle.
The (Abelian) anyonic statistic is characterized by a statistical angle
$\theta$, so that an exchange of two identical particles
gives rise to the phase factor $e^{-i\theta}$.
In Refs.~\cite{WenNiu,WDF} it was pointed out
\begin{equation}
 {\calT_x}^{-1} {\calT_y}^{-1} \calT_x \calT_y = e^{- i 2 \theta}.
\label{eq:anyon}
\end{equation}
This is because the left-hand side corresponds to worldlines of
the two quasiparticles forming two linked loops in the space-time,
as illustrated in Figs. 4 and 5 in Ref.~\cite{WenNiu}.

Let us first consider the simple case of either bosonic ($\theta=0$)
or fermionic ($\theta=\pi$) statistics, for which
$\calT_x$ and $\calT_y$ commute from eq.~(\ref{eq:anyon}).
Thus, applying $\calT_x$ does not change the eigenvalue $t_y$
of $\calT_y$ while it changes the eigenvalue of $\tilde{\calF}_x$.
Therefore, in this case, one can obtain $q$ different
eigenvalues of $\tilde{\calF}_x$ by successively applying $\calT_x$,
for each of $q$ different eigenvalues of $\calT_y$ that is
obtained by application of $\tilde{\calF}_y$.
Thus, there are at least $q^2$-fold degenerate groundstate
corresponding to the different set of eigenvalues.
In particular, when $\tilde{\calF}_x$ and $\tilde{\calF}_y$ commute,
the degeneracy deduced from the above set of algebra is $q^2$.

On the other hand,
in the Laughlin state at filling fraction $1/q$ where $q$ is odd,
the quasiparticles are known to carry the fractional charge $1/q$
($p=1$ in the previous notation)~\cite{Laughlin-FQH},
and to exhibit anyonic fractional statistics with the statistical
angle $\theta = \pi / q$.~\cite{Halperin}
In this case, because of eq.~(\ref{eq:anyon}), we obtain
\begin{equation}
 \calT_y \big( \calT_x | f_x, t_y \rangle \big)
 = t_y e^{2 \pi i / q} \big( \calT_x | f_x, t_y \rangle \big) .
\label{eq:tychange}
\end{equation}
Thus, combined with eq.~(\ref{eq:fxchange}), an application of
$\calT_x$ induces the change in both the eigenvalues
of $\tilde{\calF}_x$ and $\calT_y$ as
\begin{equation}
 (f_x, t_y) \rightarrow (f_x e^{-2\pi i /q}, t_y e^{2 \pi i/q}).
\label{eq:fxtychange}
\end{equation}
This allows the possibility that
the groundstate degeneracy on the torus to be smaller than $q^2$.
This could happen if
\begin{equation}
 \tilde{\calF}_x \tilde{\calF}_y
\sim e^{-2 \pi i /q}  \tilde{\calF}_y \tilde{\calF}_x,
\label{eq:fxfy}
\end{equation}
when acting on the groundstate subspace.
In this case,
because the application of $\tilde{\calF}_y$ induces exactly
the same change of the eigenvalues in eq.~(\ref{eq:fxtychange}),
we can generate only $q$ different set of eigenvalues.

In fact, eq.~(\ref{eq:fxfy}) is exactly what holds in the
Laughlin state.
As pointed out in Ref.~\cite{WenNiu},
because the quasiparticles and holes in the Laughlin state can
be identified with a ``vortex'' with unit flux quantum, 
the encircling process $\calT_x$ actually introduces a unit
flux quantum threading the ``hole'' of the torus, as $\calF_y$ does.
Thus eq.~(\ref{eq:fxfy}) follows.
Actually, it means that $\calT_{x,y}$ can be identified with $\calF_{y,x}$
as far as their action in the groundstate subspace is concerned.
Thus the two algebras eqs.~(\ref{eq:txfx}) and (\ref{eq:tyfy})
are indeed reduced to a single magnetic algebra,
leaving only the $q$-fold degeneracy.
On the other hand, if the statistical angle $\theta$ does not match the
fractional charge of the quasiparticles, we should have
a larger degeneracy.
When the quasiparticle statistics is non-Abelian,
the exact counting is more complicated.
Nevertheless, the minimum $q$-fold degeneracy still holds because
eq.~(\ref{eq:txfx}) is based on the fractionalized charge and
should not depend on the statistics.
The detailed discussion of the non-Abelian case is deferred
to a separate publication.

The above discussion can be generalized to a
two dimensional system on the surface with genus $g$, for which
there are $g$ pairs of intersecting elementary nontrivial cycles.
We can define the flux insertion (plus the appropriate
large gauge transformation) operator $\tilde{\calF}_c$ 
and the quasiparticle winding operator $\calT_c$
for each cycle $c$.
Picking one cycle from each pair, we have a set of
$g$ non-intersecting cycles so that the operators
for the different cycles commute.
Thus, for any (Abelian or non-Abelian) statistics of the quasiparticles,
we have $g$ independent magnetic algebras acting on the groundstate
subspace and thus the groundstate degeneracy must be at least $q^g$.
If the quasiparticle is either boson or fermion,
we can utilize $2g$ set of magnetic algebras and the
degeneracy must be at least $q^{2g}$.

The close relation between the
insertion of the unit flux quantum and trapped vortices
was emphasized previously in the {\bf Z}$_2$ gauge theory description
of a fractionalized phase.~\cite{SenthilFisher}
The adiabatic flux insertion $\tilde{\calF}_{c}$
was also used to relate topologically degenerate groundstates in
the FQHL.~\cite{WenNiu,Wen}
The present argument suggests that these structures are
rather universal in fractionalized systems.

Our argument could also be generalized to dimensions $d \neq 2$.
Although our understanding of the topological order
is still incomplete for $d \geq 3$,
our argument implies a groundstate degeneracy in
a gapped fractionalized system defined on a
geometry with a nontrivial fundamental group.
This suggests that the topological order is essential
also in $d \geq 3$.

On the other hand, the situation is quite different in $d=1$, where
the ``polyacetylene'' type fractionalization~\cite{SSH}
is known to occur in a conventional ordered phase
with a spontaneous breaking of the translation symmetry.
Our argument applied to $d=1$
just requires the groundstate on a ring to be degenerate,
as there is no higher topology.
The degeneracy can be understood as a consequence of the
spontaneous symmetry breaking of the conventional type,
rather than due to any ``topological'' order.
For $d \geq 2$, our argument reveals 
that the groundstate degeneracy indeed depends on the topology,
implying the topological order.
The present observation could help clarifying the profound
difference in the fractionalization between $d=1$ and $d \geq 2$.

Throughout this paper we have assumed the system to have a finite gap.
However, the topological order can
exist also in gapless systems~\cite{SenthilFisher},
which we have not yet analyzed.
It might be interesting to extend our argument to gapless cases.
Although the concept of the groundstate degeneracy itself becomes
subtle, the (quasi-)degenerate groundstates
may be identified separately from the gapless excitations
for example by examining the finite-size scaling carefully.

To summarize, we have derived a topological degeneracy,
which indicates the presence of a topological order,
in a general (gapful) fractionalized system in $d=2$.
The magnitude of the degeneracy is also related to the statistics
of the fractionalized quasiparticles. 
It is also notable that the simple trick of the flux insertion
together with the gauge invariance leads to the rather strong
statement, to be added
to existing applications~\cite{Laughlin,TaoWu,WenNiu,Drude,comme,
Lutt,Hastings,PV}.
Comparing with the ``momentum counting''
type applications~\cite{TaoWu,Drude,comme,Lutt,Hastings,PV}
of the flux insertion, the present argument is more powerful
in the sense that it can be applied to various topologies,
to show the degeneracy is indeed topological.
The ``momentum counting'' arguments can be applied only to
a cylinder or a torus, and thus by itself does not indicate whether
the derived degeneracy
is topological one or due to a conventional order.
On the other hand, the operator $\calT_c$ needed in
the present argument is introduced in a hand-waving way,
and thus makes the argument considerably
less rigorous than the ``momentum counting'' ones.

\medskip

This work was initiated during
2003 Summer Workshop at Aspen Center for Physics,
``Competing orders and quantum criticality''
in which both of the authors participated, and extended
while M. O. attended the 2004 ``Exotic order and criticality in
quantum matter'' program at Kavli Institute for Theoretical Physics,
UC Santa Barbara (supported in part by NSF Grant PHY99-07949). 
M. O.  thanks the participants of
the ``Exotic'' program at KITP, and Claudio Chamon
for very useful comments.
In particular, we are grateful to Gr\'{e}goire Misguich for
correcting the initial confusion on the anyon statistics.
M.O. is supported in part by Grant-in-Aid for Scientific Research ,
and a 21st Century COE Program at Tokyo Institute of Technology
``Nanometer-Scale Quantum Physics'', both from MEXT of Japan.
TS acknowledges support from NSF Grant No. DMR-0308945, 
funding from the NEC Corporation, the Alfred P. Sloan Foundation,
and an award from the The Research Corporation.

\end{document}